# VELOCITY BIAS IN MS1224+20

R. G. Carlberg[1]

Department of Astronomy, University of Toronto,
60 St. George St., Toronto, Ontario, M5S 1A1, Canada

## ABSTRACT

Velocity bias offers a single parameter ($b_v$) description of the likely segregation of mass and light in a galaxy cluster. The relation between the projected mass profile and the light profile, normalized by the virial $M/L$, is presented as a function of $b_v \equiv \sigma_\ell/\sigma_\rho$, the ratio of the RMS velocity dispersions of the galaxies and the underlying dark matter. The light distribution does not trace the mass at any radius within the cluster. At the apparent harmonic radius the gravitational mass can exceed the virial mass calculated from the light by about a factor of 2 (or more). The data for the MS1224+20 cluster, covering nearly a decade in radius, are well described by a velocity bias value of $b_v = 0.85 \pm 0.05$, comparable to the n-body estimate of $b_v = 0.8 \pm 0.1$. This particular value will remain tentative until more data are available.

*Subject headings:* galaxies: clusters of — dark matter

1. INTRODUCTION

The masses of clusters of galaxies are conveniently estimated with the virial mass, $M_v = (3\pi/2)\sigma_1^2 r_h$, where $\sigma_1$ is the line of sight velocity dispersion, and $r_h$ is the harmonic mean radius, the normalized inverse of the sum of all pair's inverse projected separations. The accuracy of this global estimator is completely dependent on the degree to which the galaxies accurately trace the cluster mass. An independent technique is to derive cluster masses from observations of the distortions of background galaxies by weak gravitational lensing. The inferred masses (Fahlman *et al.* 1994 hereafter FKSW, Bonnet, Mellier & Fort, 1994, Smail, Ellis & Fitchett 1994) substantially exceed the virial masses of the clusters as indicated by cluster galaxies. Even at the harmonic radius, where it might have been anticipated that a cluster's virial mass would be nearly equal to its gravitational mass, the excess is about a factor of 2.

It has long been recognized that the galaxies in clusters may be segregated from the cluster mass in some way (*e. g.* Limber 1959, Aarseth & Saslaw 1972, West & Richstone 1988). In principle the relation between the mass density profile and the galaxy tracer profile could take a wide variety of forms. Velocity bias is a dynamically motivated effect which predicts a specific relation between a cluster's light and mass. The velocity bias parameter, $b_v$, is defined as the ratio of the RMS velocity dispersion of cluster galaxies, $\sigma_\ell$, to the underlying cluster dark matter's velocity dispersion, $\sigma_\rho$. Theoretical values for $b_v$ are, at the moment, based on identifying plausible galaxy candidates within large N-body simulations. Most of the well resolved cluster simulations, both completely gravitational (Carlberg & Dubinski 1991, Carlberg 1994, hereafter C94) and ones that incorporate hydrodynamics (Evrard, Summers & Davis, 1992, Katz & White 1993) are compatible with a velocity bias value $b_v = 0.8 \pm 0.1$. (Here $b_v = b_v(1)$, the single particle velocity bias of C94.) The numerical value from simulations depends on the specific assumption that galaxies form in small dark halos which later aggregate to form clusters (White & Rees 1978).

Because a cluster's dark matter velocity dispersion is not directly observable a test for velocity bias must be done in other quantities. One candidate is the ratio of the mass and light profiles, where the mass profile is inferred from observations of cluster X-rays, galaxy velocities or gravitational lensing. The next section of this paper works out the practical details of velocity biased projected mass and light distributions. Section 3 compares the models to the combined lensing and dynamical observations of the MS1224+20 cluster to derive a velocity bias for this cluster. The final section reminds the reader of the caveats for this observational estimate of cluster velocity bias.

2. VELOCITY BIAS MODIFIED CLUSTER LIGHT PROFILES

In detail the dynamical state of galaxy clusters remains a research topic, but the central parts of most clusters can be considered to be in an approximate stellar dynamical equilibrium. The Jeans equation describes the balance of gravitational force against the gradient of the radial "pressure", $\nabla(\nu\sigma_v^2)/\nu$ (for an isotropic velocity ellipsoid), where $\nu$ is the number density of galaxies and $\sigma_\ell$ is their velocity dispersion. If the galaxies have a lower velocity dispersion than the dark matter then they will have a steeper density gradient to create the same stellar dynamical pressure gradient. The equilibrium assumption becomes completely invalid between 1 and 2 harmonic radii of the cluster mass beyond which the dynamics is dominated by infall and becomes strongly dependent on both $\Omega$ and local structure (Quinn, Salmon & Zurek 1986, Zurek, Quinn, & Salmon 1988, Crone, Evrard & Richstone 1994). Because velocity bias reduces the harmonic radius of the light a factor of 4 or so below that of the mass, the dynamics of the galaxies are reasonably approximated as being in a nearly equilibrium potential (unless of course the cluster is undergoing a massive merger).

To make a prediction for the $M(r)/L(r)$ ratio requires an assumption for the underlying $M(r)$ profile of clusters. A simple analytic mass profile which is a reasonable description of much of the radial range of dark matter halos found in n-body experiments (Dubinski & Carlberg 1991) is Hernquist's (1990) potential, $\phi = GM/(r+a)$. This function is a useful approximation within about twice the half mass radius, $r_{1/2} = (1+\sqrt{2})a$, for the reasons mentioned above. The velocity dispersion as a function of radius within this potential is calculated (Hernquist 1990) assuming an isotropic velocity ellipsoid, which is a first order description of the spherically averaged n-body simulation data which typically has a $\sim 20\%$, velocity anisotropy (Carlberg & Dubinski 1991, Dubinski 1992, Crone, Evrard & Richstone 1994). The surface mass density, $\Sigma_\rho$, is numerically derived from the volume mass density.

The surface light density, $\Sigma_\ell$, of a cluster is straightforward to calculate assuming that the $b_v = \sigma_\ell/\sigma_\rho$ is a constant (C94). The current n-body data, and the argument that the virialized interior of a dark halo is nearly a scale free power law, suggest that this should be a satisfactory initial approximation. The resulting equilibrium light distributions give a mean harmonic radius and RMS velocity dispersion which when combined in the virial theorem indicate a virial mass, as indicated by the luminous matter. The resulting $M/L$ of the luminosity distribution greatly underestimates the total mass of the cluster, the effect being about a factor 4 at $b_v \simeq 0.85$. The global $M/L$ value is here used to normalize the 3D luminosity profiles to mass profiles of C94, which are then projected to give the light-traces-mass profiles $\Sigma_\ell(r)$ displayed in Figure 1. Note that the $\Sigma_\ell$ is always steeper than $\Sigma_\rho$. The harmonic radius of the unbiased model is $12/\pi$ units in projection so the radius where $\Sigma_\ell$ is predicted to exceed $\Sigma_\rho$ will be 10-100$h^{-1}$ kpc, and will be quite sensitive to the details of the inner halo and mass profile of the central galaxy (often a cD which extends out to this distance).

## 3. OBSERVATIONS

The $z = 0.327$ cluster MS1224+20 (Gioia *et al.* 1990) was imaged at the Canada-France-Hawaii Telescope as described by FKSW, from which they derived mass surface densities using the Kaiser & Squires (1993) algorithm. Photometric and velocity data for the cluster were also acquired at CFHT as described by Carlberg, Yee and Ellingson (1994, hereafter CYE) as part of the CNOC cluster mapping project (Carlberg *et al.* 1994). CYE found the cluster has a velocity dispersion of 770 km s$^{-1}$ on the basis of 30 members, and a virial mass of $2.1 \times 10^{14} h^{-1} M_\odot$, giving an unremarkable $M/L_V(0) = 255 \pm 30\% h M_\odot/L_\odot$ (where the light is both redshift and evolution corrected to a $z = 0$ value). On the other hand, the FKSW weak lensing surface density map integrated out to $2.76'$ indicates a gravitational mass about 3 times the virial mass The *average* surface mass densities within an annulus (that is, $\mu(r) = 2r^{-2} \int_0^r \Sigma x dx$) are derived in FKSW (for the gravitational mass) and CYE (for the light as normalized by the virial $M/L$), and are displayed in Figure 2. The values plotted are normalized to the critical surface density of the cluster, $\Sigma_c = 7.5 \times 10^{15} M_\odot h^2$ Mpc$^{-2}$, which depends on an assumption about the redshifts of the background galaxies, discussed in FKSW. The weak lensing mass estimate is not corrected for faint cluster galaxies, which are not lensed, and therefore dilutes the distortion of the background background galaxies. Therefore the $\mu_\rho$ are likely somewhat underestimated with the center being most affected. Note also that $\mu_\rho$, as an integral quantity, has correlated errors. In Figure 2 the $\mu_\ell$ has an uncertainty of 30% from the virial $M/L$ analysis. Note that these data indicate that for the radii observed $\mu_\rho$ everywhere exceeds $\mu_\ell$. A similar excess of $\mu_\rho$ over $\mu_\ell$ is found for other clusters by Bonnet, Mellier & Fort (1994) and Smail *et al.* (1994).

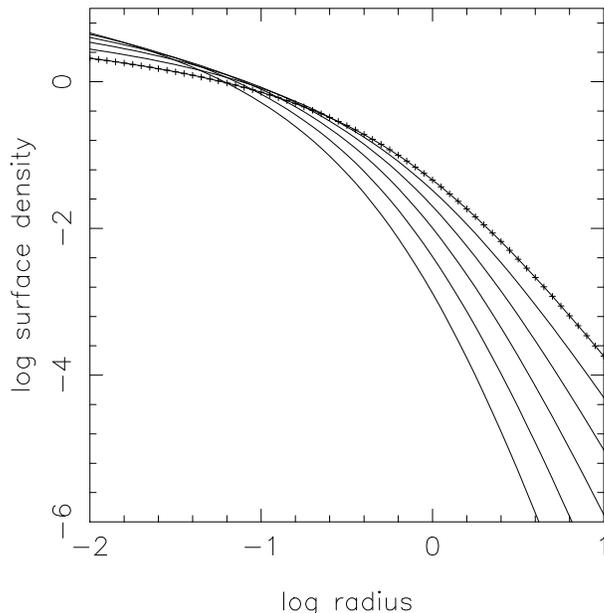

Figure 1: Velocity biased surface density profiles derived from the $\phi = GM/(r+a)$ potential. The line marked with plus signs is the unbiased mass density profile, $b_v = 1$, with the other lines are at decreasing increments of 0.05 in $b_v$.

The surface mass and light data of Figures 1 and 2 are divided to create $\mu_\rho/\mu_\ell$ which is displayed on Figure 3 as asterisks with error bars along with predictions for a range of $b_v$. The distances on the sky are normalized to a harmonic radius of $110''$. The velocity bias predictions of the ratio $\mu_\rho/\mu_\ell$ are displayed in Figure 3 as a function of the $b_v$. The circles on the curves mark the harmonic radius as would be measured from $\Sigma_\ell$ ($2/\pi$ times the 3D values, Limber and Matthews 1960). The data fall between the $b_v = 0.80$ and $0.85$ predictions, being closest to $b_v = 0.85$, which will be taken as the velocity bias indicated by the data and assign an approximate error range of $\pm 0.05$. In this display of integral data the errors from point to point are of course correlated, and the main source of error present in this data is likely to be systematic errors. Independent analysis of other clusters to derive $\mu_\rho/\mu_\ell$ is critical to asses the variations and errors of $b_v$.

## 4. DISCUSSION AND CONCLUSIONS

The aim of this paper is to demonstrate one practical measurement for the velocity bias parameter and to make an illustrative comparison to the single cluster for which all the relevant data are presently available. The result is that the cluster light and lensing mass data for the MS1224+20 cluster are well described by a velocity bias model with $b_v = 0.85 \pm 0.05$, although the result must be considered tentative until more conclusive data are available. This particular value of $b_v$, in conjunction with a Hernquist (1990) potential predicts that the equilibrium cluster mass is 3.8 times the virial mass. This cannot be used safely to infer an $\Omega$ value because clusters are subject to continuing infall and the details of the profile are not likely to be correct at large radii. The significance of a value for velocity bias is that can be used (cautiously) to convert light profiles into expected mass profiles, and is an essential first step in understanding the nature of velocity bias in the field, which affects the Cosmic Virial Theorem.

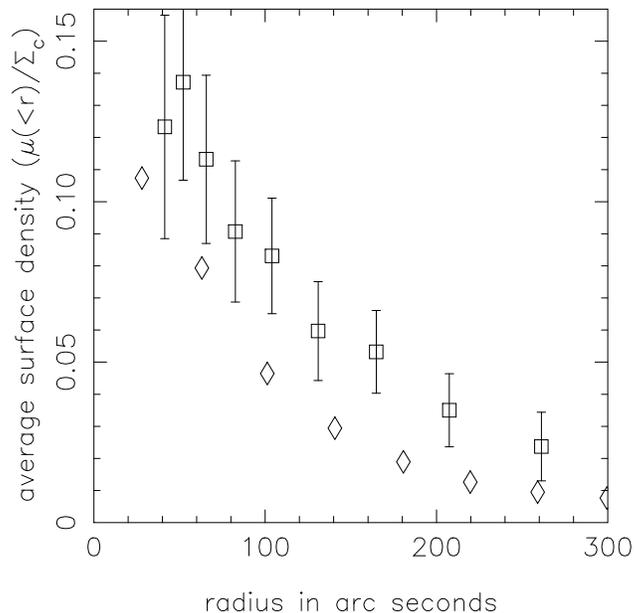

Figure 2: The surface mass density of the cluster (squares) and the light profile normalized with the virial $M/L$ ratio (diamonds). Both quantities are normalized with the critical surface density of the cluster, $7.5 \times 10^{15} h^2 M_\odot$ Mpc$^2$.

The main concern with the velocity bias value derived is that it is based on a single cluster. The virial analysis is well defined and the $M/L$ derived is not controversial, being comparable to values obtained for many low redshift clusters, although it is always important to have sufficient velocities to weed out unrelated nearby substructure and reduce random errors. The lensing observations require great care to correct for instrumental distortions in the galaxies, and, at the moment the redshifts assigned to the background galaxies are uncertain. A further complication is that structure along the line of sight and near the cluster introduce "cosmic noise" in the mass measurement. All of these sources of error were considered in the FKSW and CYE papers, and estimated to be sufficiently small that the mass ratio derived could be treated with confidence. Two other groups (Bonnet, Mellier & Fort 1994, Smail, Ellis & Fitchett 1994) have independently found similar results for other clusters. It is worth emphasizing that all groups have claimed an overall mass excess, but have not been confident that the data indicates that the mass is more extended than the light.

On the theoretical side, the detailed understanding necessary to predict a value for the velocity bias does not exist, nor are the current simulations without various numerical difficulties, relating first to numerical resolution, and secondly to the fundamental problem of galaxy formation within a simulation. However, within the context of a hierarchical gravitational instability model for structure and galaxies forming within dark halos the results should be relatively robust.

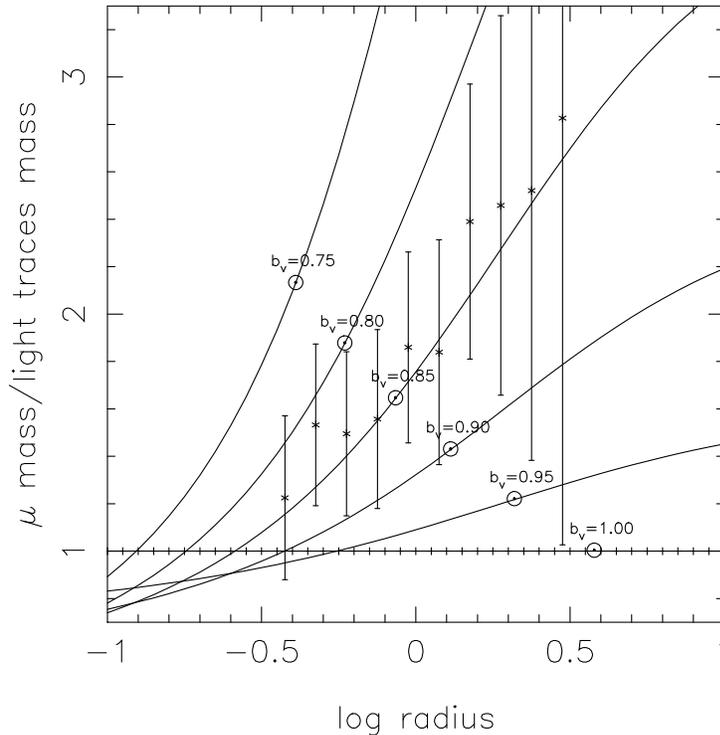

Figure 3: The ratio of the average mass profile, $\mu_\rho$, to the virial normalized light profile, $\mu_\ell M/L$, for various values of the velocity bias. The observational data for the MS1224+20 cluster is reasonably described with $b_v = 0.85$, approximately the same value as found in n-body experiments.

## ACKNOWLEDGMENTS

I thank Greg Fahlman, Nick Kaiser, Gordon Squires and David Woods and the CNOC collaboration for use of the data in preprint form. This research was supported by NSERC.